\documentclass[conference]{IEEEtran}
\usepackage[utf8]{inputenc}
\usepackage{amsmath}
\usepackage[ruled,vlined,linesnumbered]{algorithm2e}
\usepackage{booktabs}
\usepackage{graphicx}
\usepackage[utf8]{inputenc}
\usepackage{multicol}

\title{MDEA: Malware Detection with \\
Evolutionary Adversarial Learning}

\author{\IEEEauthorblockN{Xiruo Wang}
\IEEEauthorblockA{Department of Computer Science \\
The University of Texas at Austin\\
Austin, USA \\
kevinxrwang@utexas.edu}
\and
\IEEEauthorblockN{Risto Miikkulainen}
\IEEEauthorblockA{Department of Computer Science \\
The University of Texas at Austin\\
Austin, USA \\
risto@cs.utexas.edu}}
\begin{document}

\maketitle

\begin{abstract}
Malware detection have used machine learning to detect malware in programs. These 
applications take in raw or processed binary data to neural network models to classify 
as benign or malicious files. Even though this approach has proven effective against dynamic 
changes, such as encrypting, obfuscating and packing techniques, it is vulnerable to 
specific evasion attacks where that small changes in the input data cause 
misclassification at test time. This paper proposes a new approach: MDEA, an Adversarial Malware 
Detection model uses evolutionary optimization to create attack 
samples to make the network robust against evasion attacks. By retraining the model with 
the evolved malware samples, its performance improves a significant margin. 
\end{abstract}

\section{Introduction}
The high proliferation of and dependence on computing resources in daily life has greatly increased the potential of malware to harm consumers\cite{acquisti}. It is estimated that almost 
one in four computers operating in the U.S. were already infected by malware in 2008 \cite{plonk2008malicious} and according to Kaspersky Lab, up to one billion dollars was stolen 
from financial institutions worldwide due to malware attacks in 2015 \cite{kaspersky2015carbanak}. More 
recently, the notorious and widespread NotPetya ransomware attack is estimated to have 
caused \$10 billion dollars in damages worldwide. Even worse, as reported by McAfee 
Labs, the diversity of malware is still evolving in expanding areas such that in Q1 2018, 
on average, five new malware samples were generated per second \cite{cruz2018mcafee}. As a 
specific example, total coin miner malware rose by 629\% in Q1 to more than 2.9 million 
samples in 2018 \cite{cruz2018mcafee}.

As a result of the magnitude of the threat posed by malware, a great deal of research 
has been done on malware identification. At the moment there are two 
widely used approaches: dynamic analysis, which obtains features 
by monitoring program executions, and static analysis, which analyzes features of binary 
programs without running them. Intuitively, the dynamic analysis is the first choice, 
since it can provide more accurate program behavior data. However, there are many 
issues in dynamic analysis in practice. It requires a specially constructed 
running environment such as a customized Virtual Machine (VM), which is
computationally very costly when numerous samples are tested. Furthermore, in order to bypass this 
defense, some malware alter the behaviors when they are detected \cite{raffetseder_kruegel_kirda,garfinkel}. The analysis environment can also get false positive data from 
other software that are running in the same environment.

On the other hand, the static analysis methods also have disadvantages. The 
signature-based method such as API calls and N-grams provides the basis for most commercial antivirus products \cite{reddy2006n,narouei2015dllminer}. While it is widely
used, its ability to combat various encryption, polymorphism and 
obfuscation methods used by malware attackers is limited. Machine learning based malware 
classification technologies have 
been applied to malware detection \cite{rieck2011automatic,zakeri2015static}. which rely heavily on relevant domain knowledge 
to provide applicable features. This approach cannot adapt to fast
changing malware patterns nor polymorphism. 
In recent years, researchers have begun exploring a new frontier in data mining and 
machine learning known as deep learning. Deep learning techniques are now being 
leveraged in malware detection and classification tasks \cite{hardy2016dl4md, gibert2016convolutional,drew2017polymorphic, yan2018detecting}. Sparse Autoencocder (SAE), Convolutional Neural Network (CNN), and Recurrent Neural Network (RNN) models are all used to devise 
malware detection architectures. Although research thus far has provided promising results, there are still many open challenges and opportunities. First of all, due to the quick 
increase in the amount of novel malware, malicious techniques and patterns are changing 
and evolving rapidly. As a result, handling novel malware is one of the most pressing issues. In addition, in contrast to the natural language 
processing or computer vision tasks that are usually explored in deep learning tasks, 
malware byte files and assembly instructions have less understandable patterns. Therefore data prepossessing 
techniques are much more important. Furthermore, the adversarial attacks against neural 
networks, which only manipulate small portion of the input data to cause 
misclassification, has been proven to be a big vulnerability. Even 
though these types of adversarial attacks are less common on malware detection models 
because of the complexity and fragility of binary executables, it is possible to evade deep neural network for malware binary detection. For example, Kolosnjaji et al. \cite{kolosnjaji2018adversarial} trained a gradient-based model to append bytes to the overlay section of 
malware samples. Even though the model successfully evaded the deep neural network, 
both the model and the modification method are rather simple and cannot cover the 
complicated modifications real malware writers do such as modifying various sections based on domain knowledge.

In order to explore the data space more thoroughly, an action space is defined. It consists of 10 different modification methods of binary programs. An evolutionary optimization is used to search the best action sequence of a specific 
malware.

This paper proposes MDEA, an Adversarial Malware Detection model that 
combines a deep neural network with an evolutionary optimization at its training. MDEA consists of a 
convolutional neural network that classifies raw byte data from malware binaries, and an 
evolutionary optimization that modifies the malware that are detected. In contrast to 
simply appending bytes to the end of each file, an action space is defined for the evolutionary 
optimization to pick from and choose best action sequences for each malware sample. With 
the evolutionary learning, the probability that the generated input sample is classified as 
benign can be maximally increased. The new samples will then be fed into the detection 
network for retraining. The above steps represent a form of adversarial training \cite{NIPS2014_5423}.

The experiments were performed on 7371 Windows Portable Executable (PE) 
malware samples and 6917 benign PE samples. The results showed that MDEA increased the overall detection 
performance from 90\% to 93\% after the retraining process. This result shows that 
adversarial evolutionary training can improve both robustness and performance of the
malware detection network.

The rest of the paper structures as follows. Section 2 presents related work. 
Section 3 describes an overview of MDEA and the details of each 
component. Section 4 describes the experimental setup and discusses the results. Section 5 provides suggestions for 
further research on this topic and Section 6 draws conclusions.

\section{Related Work}
Malware detection and classification has been studied problem for many 
years. Notably, in 2015, the open Kaggle Contest: Microsoft Malware Classification 
Challenge (BIG 2015)\cite{kaggle2015malware} created a lot of interest in malware classification. The champion of this contest used machine learning 
with sophisticated static pattern analysis. Following that trend the goal of this paper is to leverage deep 
learning models without sophisticated feature engineering. This section briefly 
introduces related work in signature-based, learning-based \cite{yan2018detecting}, adversarial
based \cite{anderson2018learning} approaches, and evolutionary techniques.

\subsection{Signature-based Malware Detection}
Signature-based and behavior-based methods are widely used in the anti-malware 
industry and are often used to identify “known” malware.\cite{cloonan2017signatures} When an anti-malware solution provider identifies an object as malicious, its signature is added to a 
database of known malware. These repositories may contain hundreds of millions of 
signatures that identify malicious objects.  One of the major advantages of such signature-based 
malware detection is that it is thorough, it follows all conceivable execution ways of a 
given document \cite{souri_hosseini_2018}. Because it is simple to build such a system, 
signature-based malware detection has been the primary identification technique used by 
malware products. It remains the base approach used by the latest firewalls, email and 
network gateways. Therefore, much research has been done in this field. Santos et al.\cite{santos_brezo_nieves_penya_sanz_laorden_bringas_2010} created an opcode sequence-based malware detection system. Preda et al.\cite{preda_christodorescu_jha_debray_2007}
proposed a semantics-based framework for reasoning about malware detector. Fraley and Figueroa\cite{fraley_figueroa_2016} presented a unique approach leveraging 
topological examination using signature-based techniques. They also used data mining 
techniques in order to uncover and spotlight the properties of malicious files. Despite the 
widespread adoption of signature-based malware detection within the information security 
industry, malware authors can easily evade this signature-based method through techniques 
such as encryption, polymorphism, and obfuscation. Signature-based analysis is, therefore, 
poorly equipped to handle the current state of malware generation. 
\subsection{Learning-based Malware Classification}
Because of the weaknesses of signature-based malware detection, machine 
learning is a popular approach to signatureless malware detection. Many different 
malware detection approaches using machine learning technology have been proposed in 
recent years. These approaches include static analysis, which learns the statistical characteristics of malware (e.g. API calls, 
N-grams), and dynamic behavior analysis, which analyzes the behavior of a system against 
a baseline in order to determine anomalous (and possibly malicious) behavior. This paper focus on 
static analysis. In the Kaggle Microsoft Malware Contest \cite{kaggle2015malware}, the winner used many sophisticated features for their K-nearest Neighbor(KNN) model in order 
to achieve high performance. Some other machine learning techniques were also studied in 
different works. Rehman et al.\cite{rehman_khan_muhammad_lee_lv_baik_shah_awan_mehmood_2018}reverse-engineered the Android Apps to extract manifest files, and employed machine-learning 
algorithms to detect malware efficiently. They observed that SVM in case of binaries 
and KNN in case of manifest xml files are the most suitable options for robustly detecting  malware in Android devices. Yuan et al. \cite{yuan_lu_xue_2016} proposed to associate features from static analysis 
with features from dynamic analysis of Android apps and characterize malware using 
deep-learning techniques.

Unlike all the above research, in order to cut down on the necessity of 
expert analysis, MDEA will only use basic features as input to a deep learning model. Many 
different deep learning models have been proposed for malware detection. Some people 
intend to solve this problem by Long Short-Term Memory (LSTM) model \cite{yan2018detecting}, while others propose 
"malware image" that are generally constructed by treating each byte of the binary as a gray-scale pixel value, and defining an arbitrary “image width” that is used for all 
images \cite{Singh2019MalwareCU}. From all these approaches, the MalConv network\cite{raff2017malware}, 
which use raw byte embeddings as the input, achieved the highest accuracy. Therefore, this model is used as our detection model.

\subsection{Adversarial Model for Sample Generation}
Nguyen et al. \cite{nguyen2014deep} inspired later research on adversarial 
models. They found that it was easy to produce images that were unrecognizable 
to humans, but deep neural networks (DNNs) could recognize them with high confidence. 
They trained DNN on ImageNet and MNIST datasets and produced many human-unrecognizable images. Goodfellow et al.\cite{goodfellow2014generative} proposed the first generative adversarial 
nets (GANs). GANs consist of two models. One of them is a generative model, 
which captures the data distribution; the other is a discriminative model that estimates the 
probability that a sample came from the training data rather than from the generative model. 
GANs have a large potential since it can learn to mimic any data distribution. Therefore, GANs have been used in many domains 
such as computer vision and natural language processing.

Recent work in adversarial machine learning has shown that deep learning models 
for machine learning are susceptible to gradient-based attacks. Anderson\cite{anderson2018learning} proposed a more general framework based on reinforcement learning (RL) for 
attacking static portable executable (PE) anti-malware engines. They showed in experiments 
that this adversarial learning method can attack a gradient-boosted machine learning model and evade components of publicly hosted antivirus engines. Kolosnjaji et al.\cite{kolosnjaji2018adversarial} 
proposed a gradient-based attack model that is capable of evading a deep network by only 
changing few specific bytes at the end of each malware sample, while preserving its 
intrusive functionality. They were able to decrease 
the detection accuracy of the original detection model by more than 50\% . Even though this work achieved good result, 
they did not use the generated attack samples to improve the detection model.

This paper will reproduce and improve upon their methods\cite{anderson2018learning,kolosnjaji2018adversarial} with evolutionary 
learning and leverage the generated adversarial samples by retraining the deep learning model 
to further improve the accuracy.

\subsection{Evolutionary Algorithm}

Evolutionary algorithms (EAs) use mechanisms inspired by biological evolution, 
such as mutation, recombination and selection to select the best individual of a population 
to solve an optimization problem. Initially, EAs were considered as a scalable alternative to reinforcement learning\cite{salimans2017evolution}. In recent years, EAs performed much better than traditional optimizing 
method such as gradient descend in various domains \cite{real2018regularized,Podryabinkin_2019}. This advantage of EA is becoming more important in deep neural network because of the diversity and complexity it provides\cite{young_rose_karnowski_lim_patton_2015,8552944}. Martní et al.\cite{Mart2017} created an Android malware detection system with 
evolutionary strategies to leverage third-party calls to bypass the effects of
concealment strategies.  Petroski et al.\cite{such2017deep} evolved the weights of a DNN with genetic 
algorithm (GA) to perform well on hard deep RL problems, including Atari and 
humanoid locomotion.  Chen et al.\cite{chen2019} built a model to generate 
groundwater spring potential map. They utilized GA to perform a feature selection 
procedure and data mining methods for optimizing set of variables in groundwater spring 
assessments.

In all the above research, EA showed an advantage against the traditional 
optimization methods such as stochastic gradient descent and reinforcement learning. Such 
gradient-free nature makes EA less vulnerable to local minimum and easier to find general solutions in high-dimensional parameter search problems. 

EA is also highly robust and performs well when the number of time steps in an episode is 
long, where actions have long-lasting effects, and when no good value-function estimates are available\cite{salimans2017evolution}. Therefore EA well suited as the optimization method for MDEA.

\section{Model}
This section presents the design and process of the proposed MDEA model. 
At first, an overview of the model structure, the dataset 
information and the details of the detection model are presented. The definition and 
details of action space, which is used by evolutionary optimization method are discussed, and the evolutionary optimization algorithm in described.

\subsection{Structure Overview}

Overall, the malware detection process consists of two major parts (Figure 1). The first part involves preprocessing malware sample data 
and feeding these data to 1-D Convolutional network. The second part uses 
evolutionary optimization to evolve adversarial malware samples to evade the network. All 
the newly generated malware that successfully evade the detection model will be added into the training set, and the detection model is retrained. The first and the second part together form 
a loop as shown in Figure 1. During the training phase, this loop is iterated multiple times 
until the detection accuracy of the detection model converges to an acceptable level.

Since the development set is fitted multiple times during the evolutionary phase, overfitting check is performed with another small set that is never used during training.

\subsection{Dataset and Detection Model}

 The dataset consists of 14,288 PE file, and 7371 of them are malware samples, that 
were downloaded from VirusShare. The rest 6917 PE files are benign files that were 
gathered by crawling different websites. The deep neural network trained and attacked in 
this paper is the MalConv Network proposed by Raff et al.\cite{raff2017malware}. Figure 2 shows the 
detailed structure of this Network. MalConv takes in up to $k$ bytes data as input. Each byte is represented by a number A = \{0, … ,255\}. The $k$ bytes that are extracted 
from input file are padded with zeros to form a vector $x$ (if there are more than $k$ bytes in 
the file, just the first $k$ bytes without padding). Each element of vector $x$ is fed 
into a trainable embedding layer to get an embedded vector $z$ of eight elements. After this 
embedding process, one-dimensional vector $x$ becomes a matrix Z $\subseteq$ $R_dx$8.

\begin{figure}[h]
  \centering
  \includegraphics[width=\linewidth]{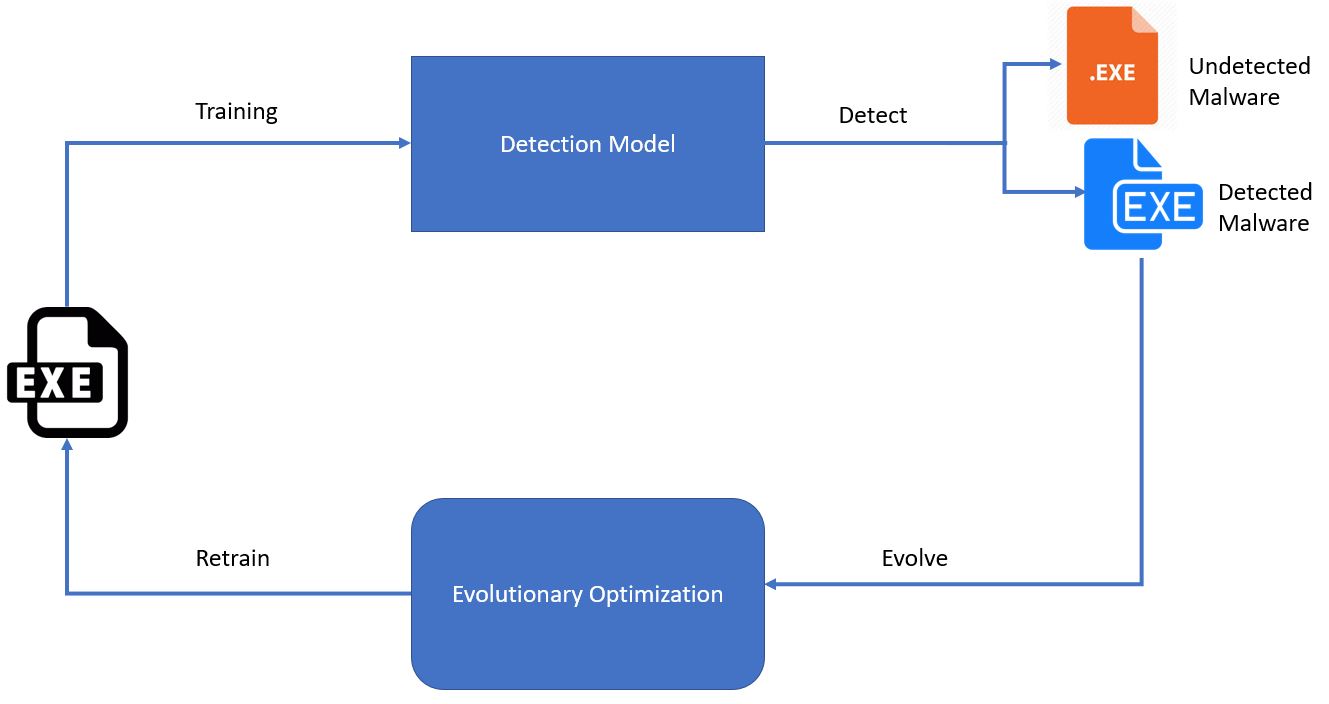}
  \caption{The overall flow of MDEA.Top: Detection Model based on the MalConv network\cite{raff2017malware}, Bottom: The Evolutionary Optimization Algorithm. In this cycle, the evolutionary method learns modification patterns for each malware to evade the detection network. The newly generated samples are then fed into detection network to improve its performance. With this cycle, MDEA not only prevents the adversarial attack against detection model, but also increases the accuracy. }
\end{figure}

This matrix Z is 
then fed into two 1-d convolutional layers. These two layers use Rectified Linear 
Unit (ReLU) and sigmoidal activation functions respectively. By combining these two 
layers with gating, the vanishing gradient problem caused by sigmoidal activation 
functions is avoided. The obtained values are fed to a temporal max-pooling layer 
followed by a fully connected layer with ReLU activation. The final classification is made based on the output of the last fully connected layer. If their output is greater than 0.5 then the 
sample is taken as a benign file, otherwise it is classified as malware.

\subsection{Action Space}

The action space is based on the PE file layout.

A PE file consists of a number of headers and sections that tell the dynamic linker 
how to map the file into memory. In general, there are three types of layouts in PE: header, 
section table and data. Header is a data structure that contains basic information on the 
executables. 
Section table describes the characteristics of each file section. 
The data layout contains the actual data related to each section. Malware usually 
modifies some of these structures to create malicious activities that are hard to detect. 
The 
stealth and sensitivity of these modifications make it harder to alter bytes 
without breaking the malware functionality. However, some of the sections are not 
important for the program to run. Some of them are even neglected by OS such as the overlay section. Based on such information the following action space is designed, inspired by Anderson et al.
\cite{anderson2018learning}.
There are 10 different actions with trainable parameters. Each action 
accepts random parameters to test evasion on a gradient-boosted decision-tree model with 
reinforcement learning.

\begin{figure}[h]
  \centering
  \includegraphics[width=\linewidth]{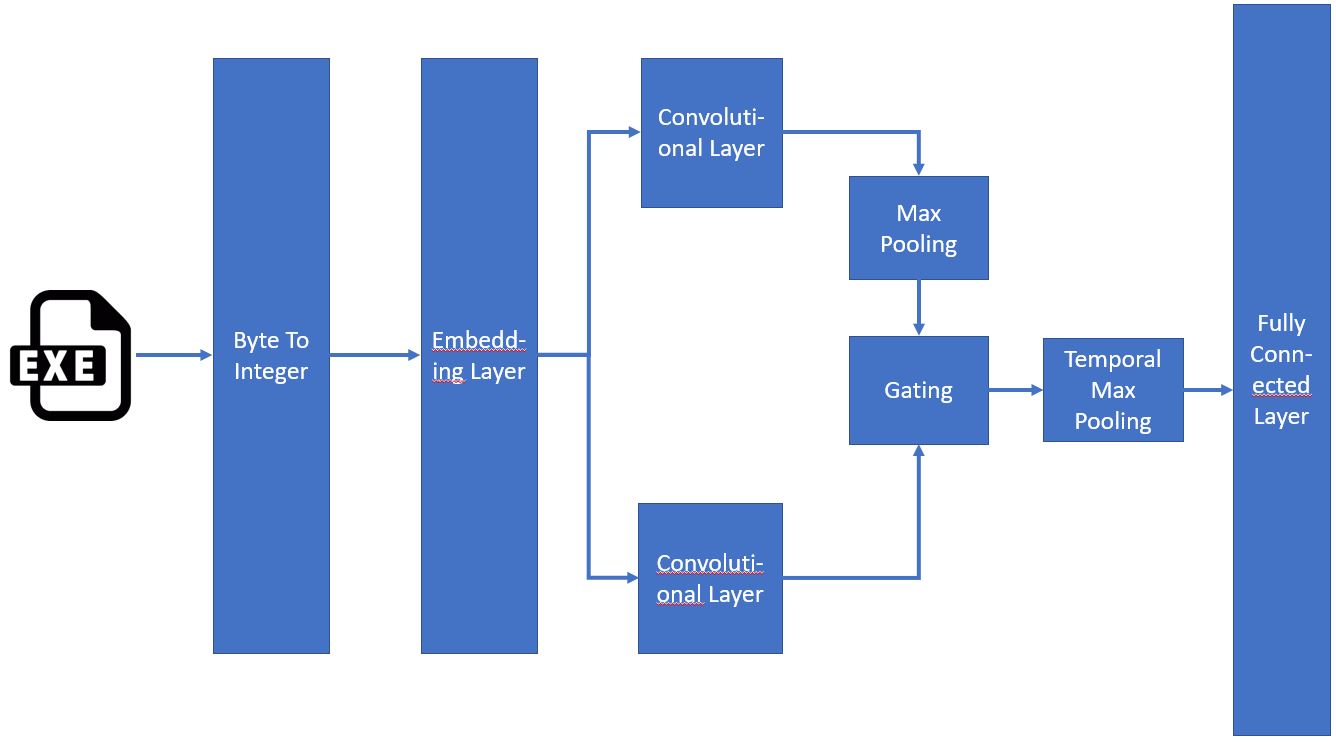}
  \caption{The details of each layer in the detection model MalConv\cite{raff2017malware}. The raw binary data is first converted to integers and then fed into an embedding layer and two convolutional layers. Then the resulting vector is fed into a fully connected layer for output. With this specific model structure, each raw byte malware sample can be processed and classified for later evolutionary optimization.}
\end{figure}

However, this randomness becomes a big issue in MDEA since the evolutionary algorithm already introduces enough generality to the problem, and more 
randomness would cause the model to not converge. Therefore, a parameter set 
is used with each action to make the model converge with acceptable time. 
 The actions are:

 \begin{itemize}
\item  Add a function to the import address table that is never used 
\item Manipulate existing section names 
\item Append bytes to extra space at the end of sections 
\item Create a new entry point 
\item Manipulate signature 
\item Manipulate debug info 
\item Pack the file 
\item Unpack the file 
\item Modify header checksum 
\item  Append bytes to overlay section. 
   \end{itemize}

Note that some of these actions such as delete a signature are not recoverable. Once the evolutionary optimization algorithm chooses to perform this action, all 
later generations of this malware will not be able to effectively perform the same 
action again. This irreversibility causes the diversity to drop drastically. This issue is addressed  
with more details in Section 4.

\subsection{Dead Species}

This section discusses the dead species problem.  There are many actions in the action space that are not reversible 
such as removing signatures and modifying checksum sections. If any of these actions are 
performed on the malware, a later generation will not be able to reverse it. Since it is 
unlikely find an optimal action sequence at the beginning of evolution, picking such 
irreversible actions drastically reduces the search space. The offspring that contain 
those actions will be stuck at a bad local optimum.

\begin{figure}[h]
  \centering
  \includegraphics[width=\linewidth]{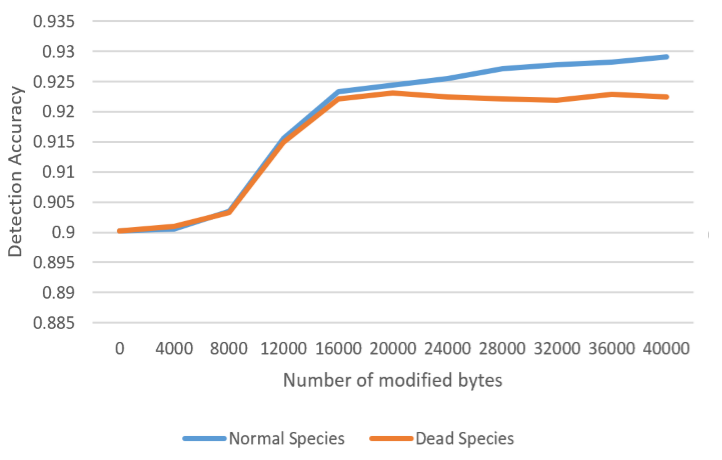}
  \caption{The performance difference between a dead species and normal species. The x-axis is the number of modified bytes. The y-axis is the detection accuracy. There is a large gap between the plots, demonstrating that irreversible actions create a limited search space and cause evolution to stagnate.}
\end{figure}

Figure 3 shows an example of such dead species. There is an obvious gap 
between the normal species and the dead species after 16,000 modified bytes. 
Investigation of these two evolution run showed that the dead 
species picked “delete-checksum” and “remove signature” actions when number of 
modified bytes was equal to 15,879. This result validates the conjecture that irreversible 
actions cause the dead species problem.

In order to solve this problem, validation weights can be introduced into the 
evolutionary optimization algorithm. These weights represent the probability of each 
action being picked. The actions that can cause dead species are assigned with a very low 
probability. This countermeasure worked well in practice and increased the average 
accuracy by 1\%. However, the validation weights only delay the occurrence of dead 
species, instead of actually solving the problem. Any individuals that pick irreversible actions lead to dead species and lose their diversity in evolution. Further possible techniques 
for dealing with this problem will be discussed in Section 5. 
\subsection{Evolutionary Optimization}

A framework called DEAP (Distributed Evolutionary Algorithms in Python) is used
to construct evolutionary optimization algorithm. It consists 
of three major parts: population initialization and evolution, binary modification, and 
population evaluation (Figure 4).

The population individual evolution part breeds new children using mutation and crossover. Mutation alters one or 
more gene values in a chromosome. There are many different types, such as shrink mutation\cite{ronco_benini_2013}, uniform mutation and boundary mutation. The “mutShuffleIndexes” method were chosen in DEAP, which shuffles the attributes of the input individual. During the 
shuffle phase, there is also a probability to replace some of the elements in those attributes 
with randomly chosen elements. In contrast, crossover combines the genetic information of two 
parents to generate new offspring.
It is one way to stochastically generate 
new solutions from an existing population. The uniform crossover method was chosen for MDEA, chooses each attribute from either parent with equal probability.
After the population is evolved, the 
action sequence will be sent to the binary modification section to produce modified 
malware.

\begin{figure}[h]
  \centering
  \includegraphics[width=\linewidth]{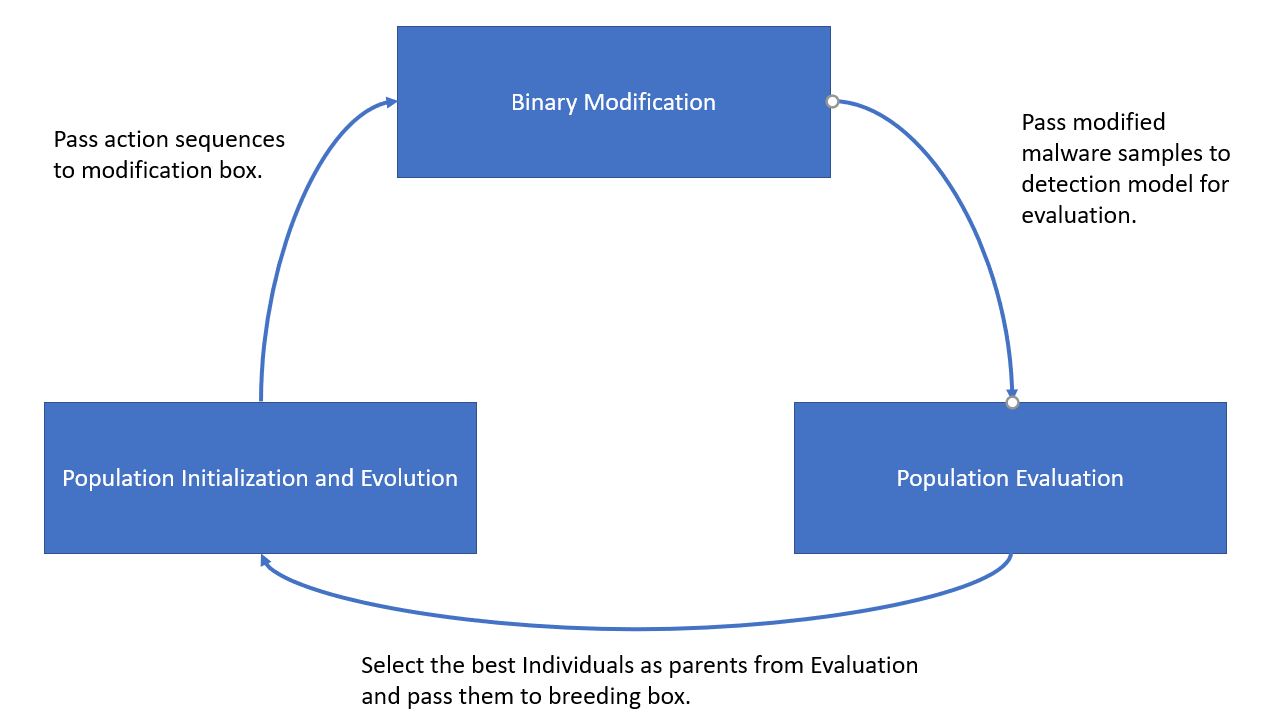}
  \caption{The evolutionary optimization cycle. The population initialization and evolution phase generates candidate action sequences. The binary modification phase modifies malware samples based on the action sequences. The population evaluation phase selects the best individual out of a population. The process then repeats for several generations until a good sequence is found. The process is described in more detail in Algorithm 1.}
\end{figure}

The modified malware will be evaluated by the detection model and the 
statistics for evolving next generation will be calculated. After the evaluation step, the 
selection method picks the best offspring as the parent for the next generation. For simplicity, the “selectBest” selection algorithm was used in MDEA.  This cycle continues until there is enough data for further training, or the 
generation limit is reached.

Let us denote a set of malware that the detection model detects as $\mbox{M}^{+}$, and the set of malware that it does not detect as $\mbox{M}^{-}$. The goal of evolution is to find a function $f$ such that $f$($m$) $\in$ $\mbox{M}^{-}$ for as many $m$ $\in$ $\mbox{M}^+$ as possible. The 
function is represented by an action sequence {($a_0$, $p_0$), ($a_1$, $p_1$), ($a_2$, $p_2$) … ($a_n$, $p_n$)} where $a_i$, $i \in$ [0..$n$], 
is an action from the action space and $p_i$ is the corresponding parameter for that action (e.g. 
for the appending overlay method, $p_i$ indicates how many bytes to append at the end).
Evolutionary optimization aims to find $f$ by crossover, mutation and selection of such sequences. Let us denote the detection model as D. For any malware sample $m$, if 
D($m$) $<$ 0.5, $m \in \mbox{M}^+$. Thus, the goal of evolutionary model is to  maximize $\sum_{i} \mbox{D}(f(m_i))$ where $m_i \in \mbox{M}^+$.
With these definitions, the overall algorithm for the evolution is shown in the Algorithm 1.

\section{Experiments}

This section will discuss the experimental setup and results. First,
the details of the datasets, their division into training and validation sets, and the hardware used to run this experiment are described. The experimental results and their 
importance are shown in terms of generations, modification sizes and random actions. 
\begin{figure}[h]
  \centering
  \includegraphics[width=\linewidth]{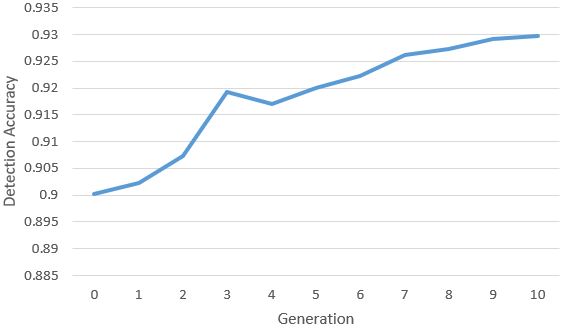}
  \caption{Detection accuracy (in y) improves with number of generations (in x), and eventually flattens out at about 10 generations.}
\end{figure}

\begin{figure}[h]
  \centering
  \includegraphics[width=\linewidth]{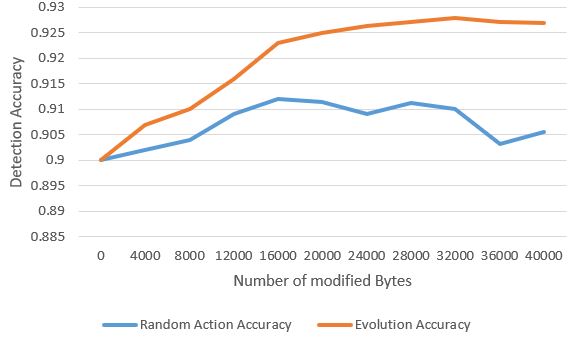}
  \caption{Detection accuracy (in y) improves with number of modified bytes (in x). The orange line represents the evolution accuracy. The blue line represents the random action accuracy. The big gap of the two lines shows
the effectiveness of evolution comparing to random actions. This gap shows that random search cannot take advantage of larger modifications, whereas evolution can.}
\end{figure}
\subsection{Experimental Setup}
To set up the experiment, 7371 malware samples were collected from VirusShare, a 
malware sample data website, and 6917 benign samples through web crawling. All these 
14,288 samples are Windows Portable Executables (PE). The dataset is generalized with comprehensive sampling and was divided into two sets 
with 9:1 ratio as training set and validation set. A extra test set is also used to check the overfitting problem. It consists of 100 malware samples that are never used during training or evolutionary phase.

Both neural network training and 
evolutionary optimization method were ran on Texas Advanced Computing Center (TACC) 
Maverick2 server with 4 Nvidia 1080-TI GPUs and 16 Intel Xeon CPUs. The training time 
for detection network was around 10 hours and the running time for evolutionary 
optimization was around 24 hours each cycle. The accuracy and the number of 
modified bytes after each cycle were recorded. 

\begin{algorithm}
\SetAlgoLined
\KwResult{Set R that contains a list of ($m$,A,P), where $m$ is a malware sample, A is an action sequence, and P is a parameter set.}
 \textbf{Parameterization:}\begin{multicols}{2}
\begin{enumerate}
 \item Malware Train Set M \item Malware $m \in$ M \item Pool of Candidates for the next Population F \item Evaluation Detection Network D \item  Current Population of (A,P,$e$) triples O\item Evaluation score $e$ \   \item  Population Size $S$\item Candidate Pool Size $L$ \item Evaluation Score Threshold $E$ \item Generation Limit $G$\item  Modified Malware $r$ \item Current Generation $g$ \item Modification Step $i$;
 \end{enumerate}
\end{multicols}
\SetArgSty{textnormal}
\For{$m$ \textbf{in} $\text{M}$}{
 $g$ = 0\;
O = RandomPickAPpairs(S)\;
\For{(A,P,$e$) \textbf{in} O}{
r = ModifyMalware(A,P,$m$)\;
e = D($r$)\;
}\
(A,P,$e$) = FindBestIndividual(O)\;
\While{$e < E$ \textbf{and} $g$ $<$ $G$}{
$i$ = 0\;
\For{$i$ $<$ $L$}{

 (A,P) = CrossOverGoodParents(O,(A,P))\;
 (A,P) = MutateGoodParents((A,P))\;
  $r$ = ModifyMalware(A,P,$m$)\;
$e$ = D($r$)\;
 F = append(F,(A,P,$e$))\;
  $i$ ++\;
 }\
 O = PickGoodIndividuals(F,$S$)\;
 (A,P,$e$) = FindBestIndividual(O)\;
 $g$ ++\;
}
R = append(R,($m$,A,P)\;
}
 \textbf{Return} R;
 \caption{Malware Sample Evolution}
\end{algorithm}
\subsection{Experimental Result}
The result graphs are shown in Figure 
5 and 6.

Figure 5 shows the detection accuracy of the detection model increases with the 
number of generations. After 10 generations of training, which is around 12 days, the accuracy 
increased from 90\% to around 93\%. Note that there is a drop at generation 4 in the graph, 
which is suspected that is caused by the dead species issues discussed in Section 4.3.

Figure 6 shows the relation between the detection accuracy and the number of 
modified bytes. There is a big jump from 8000 bytes to 12,000 bytes of the evolution line. A possible reason is that some of the sections in the PE file have specific
length requirements and when the number of modified bytes increases above the thresholds, the 
modification bytes  start to capture more malware patterns, which increases the accuracy 
by a big margin. 
In contrast, when the training set was formed by performing random actions on the malware, such a jump was not observed. The graph shows that the random action detection accuracy never rises above 0.912, compared to 0.93 of the evolution model. Therefore, evolutionary optimization can take advantage of larger malware patterns, which are more effective in training.

\section{Discussion and Future Work}
This section explains the design choices for MDEA. It also evaluates 
how well the current approach worked and what could be improved in future work.

There are several design choices. The first one is the detection 
model, i.e. the MalConv network\cite{raff2017malware}. Several malware detection methods 
were researched such as malware images\cite{Nataraj}, N-gram, K nearest neighbor 
(Kaggle 2015), and LSTM sequence model\cite{yan2018detecting}. After testing with the 
dataset, MalConv achieved the highest detection accuracy, therefore, MalConv was chosen 
as the detection network.

Another design choice was to use evolutionary algorithm (EA) as the optimization 
method to generate malware samples instead of GANs or reinforcement learning (RL). 
EA has several advantages. Existing GANs 
(GANs and its variants) suffer from training problems such as instability and mode collapse. 
EA can achieve a more stable training process. GANs usually employ a pre-defined 
adversarial objective function alternating training a generator and a discriminator\cite{wang2018evolutionary}. However, the action space cannot be simply expressed as a single adversarial 
objective function. EA solves this problem by evolving a population of different adversarial 
objective functions (different action sequences). Compared to RL, EA does not need to 
backpropagate the action weights and biases, which makes the code 2-3 times 
faster in practice. EA is also highly parallelizable compared to RL since it only requires 
individuals to communicate a few scalars between each other. Finally, EA is also more 
robust in the perspective of scaling\cite{salimans2017evolution}. It is very easy to extend the 
action space and other parameters to achieve a different learning outcome. With the above 
benefits, EA was chosen instead of GANs or RL.

The current setup of MDEA increases detection accuracy from 90\% to 93\%. Even 
though the result is promising, there are still some ways in which it can be improved. First, the evolutionary optimization still suffers from the dead species problem. The validation 
weights introduced in Section 4.3 only delay the occurrence of dead species instead of 
solving it. However, it may be possible to find an order of action sequence so that it is not necessary to reverse irreversible actions. One possible way to find such an order is to write order rules by hand; another way is to learn them from the data itself. Even if the rules are not perfect, they may help avoid dead species in many cases.

Overfitting of evolutionary optimization is another field for future 
study. The current solution is to use a unseen test set. However, 
this solution requires more data gathering time and can only be applied once. Recently, 
Feldman et al.\cite{feldman2019advantages} showed the benefits that multiple classes have on the amount of 
overfitting caused by reusing development set. With their theory and method, it is possible to extend 
the malware detection problem to multi-class malware classification problem, and the 
overfitting of development set can be reduced. The plan is to relabel each malware 
samples into different malware classes and convert the detection problem into a multi-class 
classification problem to alleviate overfitting.

Another future plan is to expand the search space for the evolutionary optimization 
algorithm by defining more modification actions. The number of actions in the action space 
is one of the key factors to ensure the diversity and generality of the evolutionary 
optimization. However, because of the fragility and sensitivity of binary EXE code, it is 
very hard to create new modification action without changing the functionality of the 
program

The size and generality of the dataset can also be improved further. Since the search 
space is constrained by the diversity of malware types, adding different kinds of malware 
into the dataset can potentially improve the optimal detection accuracy. The input size of 
the detection model can also be further increased. Currently, MDEA takes in two million bytes as the input. There are many malware samples in the dataset that have more than two 
million bytes and the extra bytes are cut off because of the length limit. The plan is to run 
MDEA with larger GPU memories, so that more data can be taken into the detection model, 
which may result in better performance. In addition, it is necessary to run an analysis in the 
future on how randomly modified malware can influence the detection model. By 
conducting this validation, the significance of the evolutionary optimization algorithm can 
be verified.

\section{Conclusion}
This paper proposed MDEA, an evolutionary adversarial malware detection 
model that combines neural networks with evolutionary optimization. An action 
space is introduced, which contains 10 different binary modification actions. The 
evolutionary algorithm evolves different action sequences by picking actions from the 
action space and then tests different action sequences against the detection model. After 
successfully evolving action sequences that bypass the detection model, all these action 
sequences are applied to corresponding malware samples to form new training set for 
the detection model. By training the network, detection accuracy increased a significant margin even with limited computing power.

These results show that deep learning-based malware detection can defend against 
adversarial attacks and accuracy can be further improved by evolutionary learning. 
Evolutionary optimization provides generality and diversity that is difficult to achieve by
other optimization algorithms.

\bibliographystyle{IEEEtran}
\bibliography{ieee} 
\end{document}